\documentclass[aps, pra, twocolumn, superscriptaddress, 10pt]{revtex4-1} 

\usepackage{amsmath,amssymb,graphicx}
\begin{document}

\title{Spin-Orbit Coupling and the Evolution of Transverse Spin}

\author{J\"org S. Eismann}
\affiliation{Max Planck Institute for the Science of Light, Staudtstr. 2, D-91058 Erlangen, Germany}
\affiliation{Institute of Optics, Information and Photonics, University Erlangen-Nuremberg, Staudtstr. 7/B2, D-91058 Erlangen, Germany}
\author{Peter Banzer}
\affiliation{Max Planck Institute for the Science of Light, Staudtstr. 2, D-91058 Erlangen, Germany}
\affiliation{Institute of Optics, Information and Photonics, University Erlangen-Nuremberg, Staudtstr. 7/B2, D-91058 Erlangen, Germany}
\author{Martin Neugebauer}
\email[]{martin.neugebauer@mpl.mpg.de}
\homepage[]{http://www.mpl.mpg.de/}
\affiliation{Max Planck Institute for the Science of Light, Staudtstr. 2, D-91058 Erlangen, Germany}
\affiliation{Institute of Optics, Information and Photonics, University Erlangen-Nuremberg, Staudtstr. 7/B2, D-91058 Erlangen, Germany}

\begin{abstract}
We investigate the evolution of transverse spin in tightly focused circularly polarized beams of light, where spin-orbit coupling causes a local rotation of the polarization ellipses upon propagation through the focal volume. The effect can be explained as a relative Gouy-phase shift between the circularly polarized transverse field and the longitudinal field carrying orbital angular momentum. The corresponding rotation of the electric transverse spin density is observed experimentally by utilizing a recently developed reconstruction scheme, which relies on transverse-spin-dependent directional scattering of a nano-probe. 
\end{abstract}

\maketitle 

\section{Introduction}
The investigation of spin-orbit interactions of light has become an integral field in modern optics, with a huge variety of related effects being relevant for a manifold of applications~\cite{Bliokh2015}. Spin-orbit coupling plays an important role in the design of spin-dependent meta-surfaces~\cite{Shitrit2013a,Karimi2014,Ling2015}, liquid crystal mode converters~\cite{Marrucci2006,Cardano2013}, and directional waveguide and plasmon couplers~\cite{Lin2013,Rodriguez-Fortuno2013,Neugebauer2014}, etc. Furthermore, it is of relevance in the field of super-resolution microscopy in the context of proper depletion beams~\cite{Rittweger2009,Hao2010}, and in optical manipulation experiments~\cite{Adachi2007,Zhao2007}.

The spin-orbit coupling also occurs naturally when a circularly polarized beam is focused~\cite{Bliokh2015}. The arising orbital angular momentum can thereby be described by a geometric Berry-phase effect~\cite{Bhandari1997}, where the longitudinal component of the field accumulates a phase of $2\pi$ for one trip around the optical axis~\cite{Bomzon2007,Bliokh2015}. The corresponding focal field distributions and their properties regarding spin and orbital angular momentum have been investigated in various works over the last decade~\cite{Bliokh2015,Nieminen2008,Bliokh2015a,Sugic2018}.

In this Letter, we report on a novel effect caused by spin-orbit interactions, which links the three-dimensional distribution of the spin density (SD) to the orbital angular momentum of the beam. Again, the effect occurs when an initially circularly polarized collimated beam of light---for simplicity, we consider a Gaussian beam profile---is tightly focused. Because of the aforementioned spin-orbit coupling, the focused beam carries not only spin, but also orbital angular momentum, which arises in the form of a phase vortex of the longitudinal field component~\cite{Zhao2007,Bomzon2007}. The superposition of the longitudinal field and the circularly polarized transverse field results in a tilted polarization ellipse offside the optical axis. Consequently, the corresponding SD features transverse components with respect to the propagation direction of the beam (optical axis). The actual local orientation of the spin depends on the relative phase between the longitudinal and transverse fields, which changes upon propagation~\cite{Pang2018}. As we will show later on, this causes the transverse components of the SD to rotate while traversing the focal region.

In the following, we start by describing the aforementioned effect as a Gouy-phase-dependent interaction of longitudinal and transverse fields~\cite{Gouy1890}. For this we elaborate on a simplified theoretical model in the framework of an extended paraxial approximation considering also longitudinal field components. 
To demonstrate the rotation of the transverse SD experimentally, we use a recently developed scheme for measuring transversely spinning fields in tightly focused light beams~\cite{Neugebauer2015,Neugebauer2018}. Finally, we compare our results with numerical calculations and discuss the possible implications of the effect on future works. 

\section{Theoretical Model}
We begin with a simplified paraxial description of a time-harmonic circularly polarized Gaussian beam. Utilizing the complex beam parameter $q\left(z\right)=z-\imath z_{R}$, where $z_{R}$ is the Rayleigh range of the beam, the transverse field components are described by~\cite{Zhan2009}
\begin{align}\label{eqn:Gauss_xy}
\left(\begin{matrix}
E_{x}\\
E_{y}
\end{matrix}\right)
= \left(\begin{matrix}
1\\
\pm \imath
\end{matrix}\right)
\frac{u_{0}}{q\left(z\right)}e^{\frac{ik\rho^2}{2q\left(z\right)}+ikz}=
\boldsymbol{\sigma}^{\pm}
u\left(\mathbf{r}\right)
\text{,}
\end{align}
with radius $\rho=\sqrt{x^2+y^2}$, $u_{0}$ a complex amplitude, and the wave number $k$. The sign of the polarization vector $\boldsymbol{\sigma}^{\pm}$ indicates right- or left-handed circular polarization.
However, the field distribution as described by Eq.~\eqref{eqn:Gauss_xy} does not fulfill Gauss's law in vacuum, $\mathbf{\nabla}\mathbf{E}=0$, which requires an additional longitudinal field component. Within the paraxial approximation, the missing component can be calculated using~\cite{Erikson1994a}
\begin{align}\label{eqn:Gauss_z}
E_{z}^{\pm}=\frac{\imath}{k}\left(\frac{\partial E_{x}}{\partial x}+\frac{\partial E_{y}}{\partial y}\right)=-\frac{u\left(\mathbf{r}\right)}{q\left(z\right)}\left(x\pm\imath y\right)
\text{.}
\end{align}
For a more intuitive description of the distribution of $E_{z}$, we rewrite Eq.~\eqref{eqn:Gauss_z} using the azimuth $\phi=\operatorname{arg}\left(x + \imath y\right)$ and the Gouy-phase $\eta\left(z\right)=\tan^{-1}\left(z/z_{R}\right)$:
\begin{align}\label{eqn:Gauss_z_LG}
E_{z}^{\pm}=-\rho\frac{\imath u\left(\mathbf{r}\right)}{\sqrt{z^{2}+z_{R}^{2}}}e^{\pm\imath\phi-\imath\eta\left(z\right)}
\text{.}
\end{align}
As we can see, $E_{z}$ is represented by a first order Laguerre-Gaussian mode with radial mode index $0$ and azimuthal mode index $\pm 1$ (the sign depends on the handedness of the incoming circularly polarized beam). Thus, the longitudinal field exhibits the helical phase-front typically associated with the occurrence of orbital angular momentum~\cite{Andrews2013}. Furthermore, in comparison to the transverse field (zero order Laguerre-Gaussian mode), $E_{z}$ exhibits an additional Gouy-phase factor~\cite{Andrews2013}. Therefore, the relative phase between longitudinal and transverse fields changes upon propagation, which consequently affects the three-dimensional polarization state~\cite{Pang2013}.

Here, we take a closer look at the evolution of the SD of the electric field $\mathbf{s}_{E}$, which describes the local orientation and sense of the spinning axis of the three-dimensional polarization ellipse~\cite{Bliokh2015,Aiello2015,Bauer2016}. For our paraxial model, we result in: 
\begin{equation}\label{eqn:sE}
\mathbf{s}_{E}^{\pm}=\frac{\epsilon_{0}\operatorname{Im}\left[\mathbf{E}^{*}\times\mathbf{E}\right]}{4\omega}
=\left(\begin{matrix}
-\frac{\rho\sin\left[\phi \mp\eta\left(z\right)\right]}{\sqrt{z^{2}+z_{R}^{2}}}\\
\frac{\rho\cos\left[\phi \mp\eta\left(z\right)\right]}{\sqrt{z^{2}+z_{R}^{2}}}\\
\pm 1
\end{matrix}\right) \frac{\epsilon_{0}\left|u\left(\mathbf{r}\right)\right|^{2}}{2\omega}\text{.}
\end{equation}
The longitudinal component of the SD has exactly the same Gaussian distribution as the electric field intensity distributions $\left|E_{x}\right|^{2}=\left|E_{y}\right|^{2}\propto s_{E}^{z}$ whereby the sign indicates right- and left-handed circular polarization. Most importantly, it is shape-invariant upon propagation. In contrast, the shapes of the transverse SD components depend on the Gouy phase and, as a consequence, also on $z$. In particular, the distributions of $s_{E}^{x}$ and $s_{E}^{y}$ rotate upon propagation along the $z$-axis. From $z=-\infty$ to $z=\infty$, the SD vector undergoes one half-twist around the $z$-axis. The rotation direction depends on the handedness of the incoming circular polarization. 

\begin{figure}[htbp]
\centerline{\includegraphics[width=1\columnwidth]{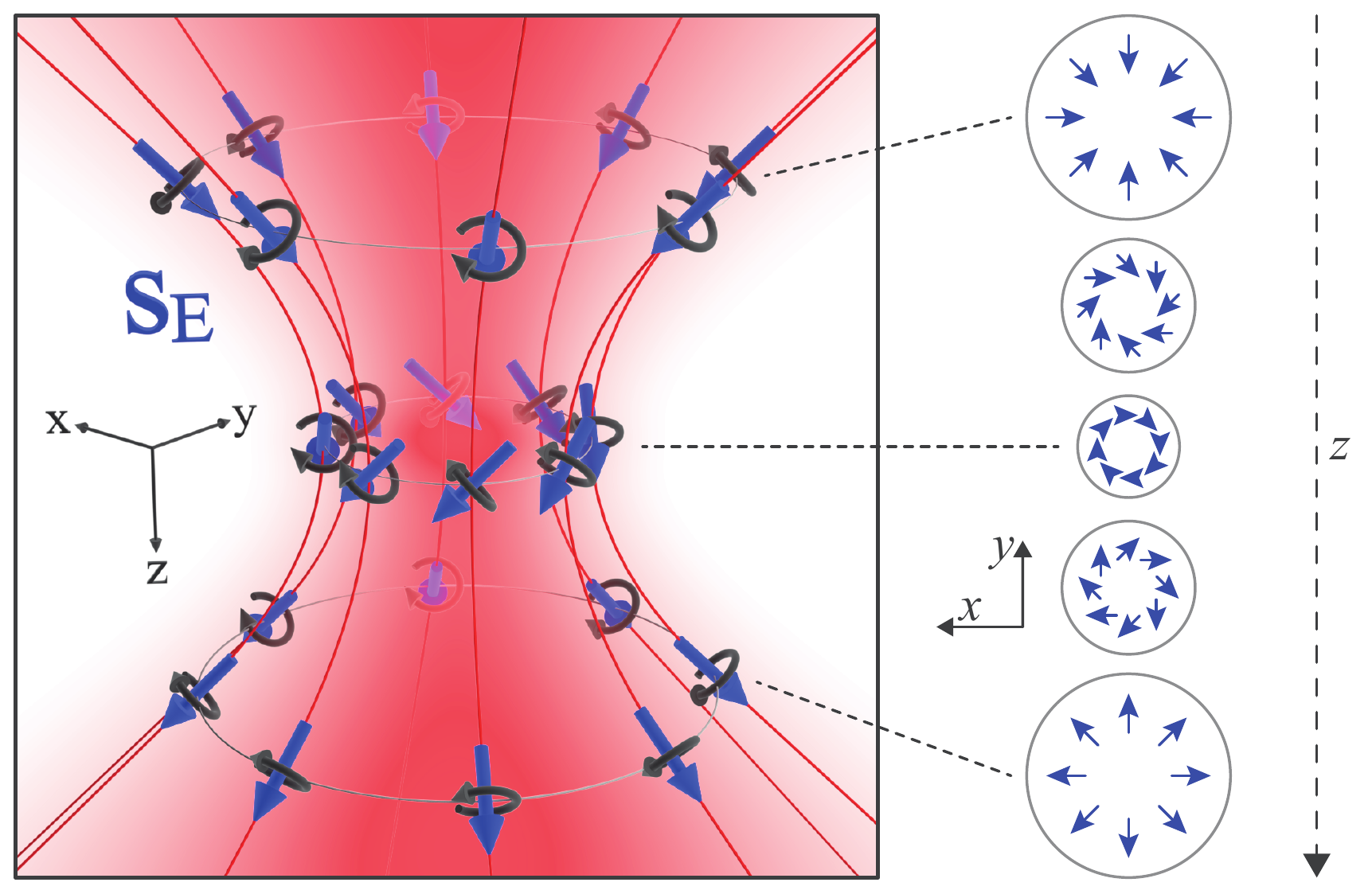}}
\caption{\label{fig_sketch}Illustration of the electric spin density $\mathbf{s}_{E}$ distribution of a tightly focused right-handed circularly polarized beam. The beam (red) propagates top-down, with the blue vectors indicating the local orientation of $\mathbf{s}_{E}$ and the black vectors corresponding to the polarization ellipse of the electric field $\mathbf{E}$. The sketches on the right side represent the transverse spin density $\mathbf{s}_{E}^{\bot}$ for different planes of observation.}
\end{figure}

For a more intuitive understanding, the effect is visualized in Fig.~\ref{fig_sketch}, where we consider a right-handed circularly polarized beam propagating top-down, with the local spin density marked as blue vectors and the corresponding orientation and spinning direction of the electric field indicated by black arrows. In the far-field of the upper half-space ($z<0$), the spin points towards the geometrical focus. A projection onto the $x$-$y$-plane reveals that the transverse SD, $\mathbf{s}_{E}^{\bot}=s_{E}^{x}\mathbf{e}_{x}+s_{E}^{y}\mathbf{e}_{y}$, is pointing towards the optical axis (see top right inset). Upon propagation, the relative phase between the transverse and longitudinal field components changes, which results in a rotation of the transverse SD. In the focal plane, $\mathbf{s}_{E}^{\bot}$ exhibits a purely azimuthal distribution (see central inset). Below the focal plane ($z>0$), the rotation continues, finally reaching a radial distribution pointing away from the optical axis in the far-field (lowest inset).

In order to confirm the half-twist of the spin density, we elaborate on this phenomenon with an experimental demonstration and numerical calculations. 
  
\section{Experimental observation}
\begin{figure*}[htbp]
\centerline{\includegraphics[width=2\columnwidth]{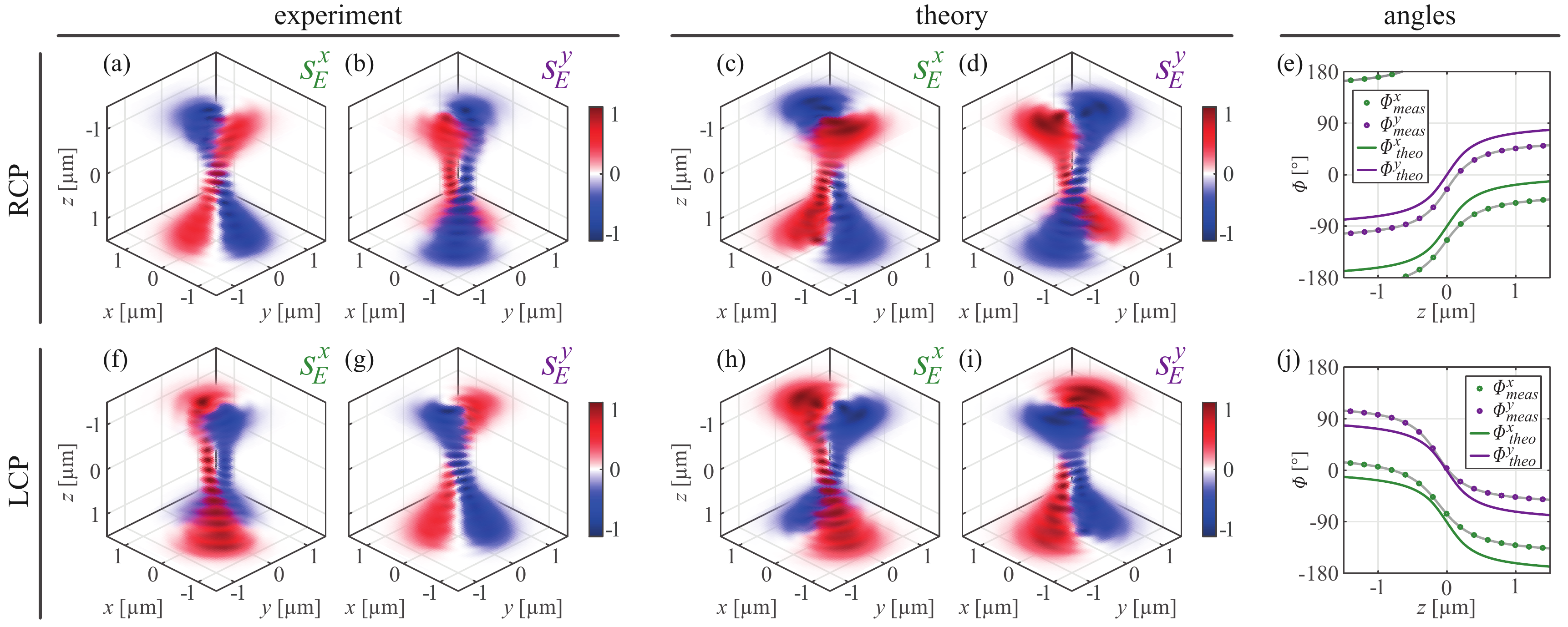}}
\caption{\label{fig_result}Experimental and theoretical results. (a)-(d) The experimentally measured and theoretically calculated transverse SD components $s_{E}^{x}$ and $s_{E}^{y}$ in the focal volume of a tightly focused right-handed circularly polarized (RCP) beam. For each plane of observation along the $z$-axis, the transverse SD is normalized to its maximum amplitude for better visibility. (e) Experimental (green and purple circles) and theoretical (green and purple lines) rotation angles of the transverse SD distributions calculated from the distributions in (a)-(d). The Gouy-phase factor is fitted to the experimental data (gray lines). (f)-(j) Similar to (a)-(e), but for a left-handed circularly polarized (LCP) beam.}
\end{figure*}

First, we investigate the rotation of the SD experimentally. Since the strength of the transverse spin depends on the lateral confinement~\cite{Aiello2015}, we investigate a circularly polarized beam (wavelength $\lambda=532\text{\,nm}$) tightly focused by a high numerical aperture microscope objective ($\text{NA}=0.9$, pupil filling factor $\approx 0.8$). The beam impinges onto a dipole-like gold nano-sphere (radius $\approx 40\text{\,nm}$) sitting on a glass substrate, which is scanned through a focal volume of $\approx 3\times3\times3 \,\mu \text{m}^3$, where we use a step size of $30\text{\,nm}$ in lateral directions ($x$ and $y$) and steps of $200\text{\,nm}$ along the propagation direction ($z$). For each position, the light scattered into the glass half space is collected with an index-matched immersion-type objective. 
The directionality of the scattered light into the substrate allows for determining the transverse SD of the excitation field at the particle position~\cite{Neugebauer2015,Neugebauer2018}. 
This is due to the fact that the directional scattering is a direct consequence of the spinning electric dipole induced in the particle by the locally transverse components of the SD~\cite{Aiello2015}. 
Finally, we assemble the scanning results to three-dimensional representations of the transverse SD components $s_{E}^{x}$ and $s_{E}^{y}$. The measurement results for right- and left-handed circularly polarized incoming beams are depicted in Figs.~\ref{fig_result}(a)-\ref{fig_result}(b) and Figs.~\ref{fig_result}(f)-\ref{fig_result}(g). 
As predicted by the paraxial model and Eq.~\eqref{eqn:sE}, the distributions of the transverse SD components rotate upon propagation, with the rotation direction depending on the handedness of the incoming circular polarization. This verifies the coupling of the transverse SD distribution and the longitudinal orbital angular momentum. For a quantitative comparison, we calculate the corresponding transverse SD distributions using vectorial diffraction theory~\cite{Richards1959,Novotny2006}, where we use the same parameters as in the experiment and consider the beam to be in free-space (effects of the glass substrate on the field distributions are not taken into account). The theoretical results are shown in Figs.~\ref{fig_result}(c)-\ref{fig_result}(d) and Figs.~\ref{fig_result}(h)-\ref{fig_result}(i). All four theoretical distributions are in very good agreement with their corresponding experimental counterparts. Also the rotation of the transverse spin densities predicted by the simplified paraxial model is confirmed by the vectorial diffraction theory. As a next step, we determine the rotation angles of the distributions of $s_{E}^{x}$ and $s_{E}^{y}$. For this purpose, we calculate the centroids of the positive and negative parts of the respective SD component for each $x$-$y$-plane of observation along the $z$-axis, and define the angle between the connection line of both centroids and the $x$-axis as rotation angle $\phi\left(z\right)$. The theoretical (solid lines) and experimental (circles) rotation angles are plotted in Figs.~\ref{fig_result}(e)~and~\ref{fig_result}(j), where the green and purple colors correspond to $s_{E}^{x}$ and $s_{E}^{y}$, respectively. For the right-handed circularly polarized beam the experimentally measured rotation angles exhibit a negative angular offset with respect to the theoretical curves, while for the left-handed circularly polarized beam the offset is positive. 
This spin-dependent offset is caused by the interference of the incoming beam and the light reflected by the glass substrate, which is not considered in the theoretical treatment. Still, the rotation angles follow a modified inverse tangent function, $\phi\left(z\right)=\pm\tan^{-1}\left[\left(z+z_{o}\right)/z_{R}\right]+\phi_{o}$, with $z_{o}$ the offset along the $z$-axis and $\phi_{o}$ the angular offset. The fits, which overlap very well with the experimental data, are plotted as gray lines. This verifies that even in the tight focusing regime, we can use the relative Gouy-phase factor between longitudinal and transverse fields derived from the paraxial model as a qualitative explanation for the half-twist of the SD.

\section{Discussion}
In conclusion, we observed the rotation of the transverse SD upon propagation in tightly focused circularly polarized beams. The effect can be explained by a difference in mode orders and, therefore, a relative Gouy-phase between the transverse field components and the longitudinal field, which carries orbital angular momentum due to spin-orbit coupling. 
The theoretical description of this effect is conceptually related to similar polarization interference effect, where two orthogonally polarized beams with different mode orders interfere resulting in a changing two-dimensional polarization distribution upon propagation due to different Gouy-phase shifts~\cite{Cardano2013,Beckley2010,Neugebauer2016a}. In our case, the effect is caused by spin-orbit interaction and occurs for the transverse SD components representing a three-dimensional polarization parameter~\cite{Setala2002}. In this regard, the measurement of the rotation of the transverse spin density can also be interpreted as an experimental demonstration of the non-separability of three-dimensional fields~\cite{Holleczek2011,Qian2011}, and the generation of orbital angular momentum by tight focusing~\cite{Bliokh2015,Nieminen2008,Bliokh2015a}. 

The evolution of three-dimensional polarization states upon propagation might find application in novel polarization based metrology approaches, where the local polarization state entails information on the position of a scatterer relative to an excitation field~\cite{Berg-johansen2015,Xi2016,Bag2018a}. By utilizing a more complex input field distribution, it is possible to tailor the rotation of the spin density for a given axis in space~\cite{Pang2018}, which might facilitate the practical implementation of such position sensing techniques and spin-based directional coupling experiments~\cite{Rodriguez-Fortuno2013,Neugebauer2014}. Finally, the notion of a position dependent orientation of the spin density in tightly focused circularly polarized beams can be relevant for optical manipulation experiments, with the local spin exerting a torque or a lateral force on nano-particles~\cite{Canaguier-Durand2013,Rodriguez-Fortuno2015}.

\section*{Acknowledgments}
We gratefully acknowledge discussions with Sergey Nechayev and Gerd Leuchs. This project has received funding from the European Union's Horizon 2020 research and innovation programme under the Future
and Emerging Technologies Open grant agreement Super-pixels No 829116.

\bibliography{bib}


\end{document}